\documentclass[aps,preprint,showpacs,epsfig,epsf]{revtex4}
\usepackage{graphicx}
\usepackage{amssymb,amsbsy}
\usepackage{amsmath}

\begin{document}
\title { Hysteresis in a Magnetic Bead and its Applications } 
\author{Vanchna Singh and Varsha Banerjee}

\affiliation{Department of Physics, Indian Institute of Technology,
Hauz Khas, New Delhi 110016, INDIA.}

\begin{abstract}
We study hysteresis in a micron-sized bead: a non-magnetic matrix embedded with superparamagnetic nanoparticles. 
These hold tremendous promise in therapeutic applications as heat generating machines. The theoretical 
formulation uses a mean-field theory to account for dipolar interactions between the supermoments. The study 
enables manipulation of heat dissipation by a compatible selection of commercially 
available beads and the frequency $f$ and amplitude $h_{o}$ of the applied oscillating field in the labortory. 
We also introduce the possibility of utilizing return point memory for gradual heating of a local region.
\end{abstract}
\maketitle

The use of small magnetic particles for diagnostic treatments is an emerging area of research. They provide several 
exciting opportunities for site-specific drug delivery, improving the quality of magnetic resonance imaging, 
manipulation of cell membranes, etc. \cite{qpankh,spuru,cberry}. The main reason behind this applicability is 
the ease with which they can be detected and directed by the application of an external magnetic field. They can 
be prepared in varying sizes, from nanometers to micrometers, comparable to the dimensions of biological entities
such as genes, proteins, viruses and cells. When coated with relevant compounds, the particles can be made to 
bind to these entities thereby providing a controllable means of tagging them. The nanometer-sized particles are 
usually single-domain and superparamagnetic (SPM) in nature. Their relaxation behavior is characterized by Neel 
and Brownian relaxation times which are strongly size-dependent \cite{rosen,vsingh}. The micron-sized particles, 
commonly called beads, are usually SPM particles packed in a non-magnetic matrix or a porous polymer with SPM 
particles precipitated in the pores \cite{mhans,qpankh}. The micron-sized beads have several advantages over 
their nanometer-sized counterparts: they are easy to observe and manoeuvre and can be bound to cells due to 
their compatible dimensions.

When sufficiently large in number, the supermoments inside the bead start interacting via dipole-dipole coupling.
They are no longer paramagnetic due to the appearance of a dipolar field. The presence of a permanent dipole moment 
opens up surprising avenues in biomedicine. An important non-equilibrium property exhibited by magnetic systems is 
that of hysteresis. Typically, when an oscillating magnetic field is applied, the response of the system is delayed 
leading to hysteresis. The magnitude of hysteresis is determined by the competition between experimental time scales 
(measured by the inverse frequency of the applied field) and the relaxation time scale of the magnetic moment. The 
area of the hysteresis loop is a measure of the heat dissipated in the system. The phenomenon of hysteresis can 
hence be advantageously used to heat magnetic beads positioned at a given site in the body, perhaps the 
site of malignancy, to selectively warm it! This effect, referred to as hyperthermia in the medical literature, makes 
it possible to destroy cells or introduce a modest rise in temperature to increase the efficiency of chemotherapy.
While the possibility of using magnetic beads for hyperthermia holds tremendous promise, its successful application
is a challenge as yet. The reason lies in the several complex questions which need to be answered in this context.
For instance, what are the clinically acceptable values of the field amplitude and frequency of the applied 
oscillating field to attain maximal heating? Or what should be the heat deposition rate to sustain tissue 
temperatures of $42^{o}$ $C$ or more? What is the adequate number of particles to be delivered at the 
specific site for achieving the required rise in temperature? It is therefore of great importance to understand 
heat generation and the role of tunable parameters for controlled heating of magnetic beads 
delivered on malignant cells.

In this letter we first motivate a mean field theory to incorporate the dipolar interactions between the magnetic 
nanoparticles embedded in the bead for the case of large uniaxial anisotropy {\it vis-a-vis} the thermal energy.  
The relaxation dynamics can then be described in terms of a two-state rate theory \cite{sdgupta1}. We study the 
role of volume fraction of SPM particles in the bead and the amplitude and frequency of the applied 
field on the hysteresis loop area. We quantify the dependence of loop area on the above parameters in the form of 
scaling laws. We also discuss the applicability of return point memory for progressive heating of a local region.

In the case of uniaxial anisotropy (in the z-direction say), the magnetic energy of a single-domain, 
SPM particle is given by \cite{sdgupta1}
\begin{equation}
\label{anis}
E = VK\sin^{2}\Phi,
\end{equation}
where $V$ is the magnetic volume of a particle, $K$ is the effective magnetic anisotropy constant and $ \Phi $ 
is the angle between the z-axis and direction of the ``super'' magnetic moment of the single-domain particle. 
Minimum energy occurs at $\Phi $ = 0 and $\pi$ and these angles define the two equilibrium orientations of the
magnetic moment. If $VK\gg k_{B}T$, where $k_{B}$ is the Boltzmann constant and $T$  the absolute 
temperature, the magnetic moment is mostly locked in two minimum energy orientations resulting in
the so-called Ising limit. In this limit, $\Phi(t)$ may be viewed as a dichotomic Markov process in which it
jumps at random between the angles 0 and $\pi$. The jump rate governed by the  Arrhenius-Kramer formula
\begin{equation}
\label{rateconst}
\lambda_{0\rightarrow\pi}  =\lambda_{\pi\rightarrow0}= \lambda_{o}\mbox{exp}\left(-\frac{VK}{k_{B}T}\right),
\end{equation}
where $\lambda_{o}$ is the ``attempt'' frequency. The reciprocal of the jump rate is the Neel relaxation 
time $\tau_{N}$.

In order to incorporate the effects of dipolar interactions, it is essential to add the contribution due to the 
``dipolar Hamiltonian'' to the magnetic energy of the particle defined by Eq.~(\ref{anis}). In the limit of large 
anisotropy, this contibution is given by \cite{abrag,schakra}:
\begin{equation}
\label{dip1}
\mathcal{H}_{d} = \sum_{i,j}\gamma_{i}\gamma_{j}\hbar^{2}\frac{(1-3\mbox{cos}^{2}\theta_{ij})}{\mid r_{ij}\mid^{3}}
m_{zi}m_{zj},
\end{equation}
where $\gamma_{i}$ and $\gamma_{j}$ are the gyromagnetic ratio of $i^{th}$ and $j^{th}$ particle respectively,
$\vec{r_{ij}}$ is the distance between the two particles, $\theta_{ij}$ is the angle between $\vec{r_{ij}}$
and the anisotropy ($z$) axis and $m_{zi}$ is the magnetic moment of the $i$th nanoparticle along the
anisotropy axis. Note that the angular dependence $\left(1-3 \mbox{cos}^{2}\theta_{ij}\right)$ implies that
the interaction can change sign thereby switching from ferromagnetic for angles close to the anisotropy axis
to antiferromagnetic for intermediate angles $(55^{o}\le\theta_{ij}\le 125^{o})$. Since $m$ is proportional to
the volume of the particle, Eq.~(\ref{dip1}) can be rewritten as
\begin{equation}
\label{dip2}
\mathcal{H}_{d} = \gamma\hbar^{2}\mu^{2}V^{2}\sum_{j}\gamma_{j}
       \frac{\left(1-3\mbox{cos}^{2}\theta_{ij}\right)}{\mid\vec{r_{ij}}\mid^{3}}
                \mbox{cos}\Phi_{i}\mbox{cos}\Phi_{j},
\end{equation}
where $\mu$ is the magnetic moment per unit volume and $\Phi$ is as defined in Eq.~(\ref{anis}).

A mean-field approach is invoked to treat the complicated interaction in Eq.~(\ref{dip1}) \cite{schakra}. 
Each particle is visualized to be experiencing an effective local magnetic field $H$ due to the surrounding 
(magnetic) medium. Further, the Ising limit allows for the replacement of cos$\Phi$ by a two-state variable 
$\sigma$. These simplifications yield the mean-field dipolar Hamiltonian:
\begin{eqnarray}
\label{hsc}
\mathcal{H}_{d}^{MF} &=& \mu V \sigma H,
\end{eqnarray}
with the field $H$ evaluated self-consistently:
\begin{eqnarray}
 H &=&  \mu \Lambda V \langle \sigma \rangle = \mu \Lambda V \mbox{tanh}\left(\frac{\mu V H}{k_{B}T}\right).	
\end{eqnarray}
The parameter $\Lambda$ contains all the constants. It fluctuates in sign and has a magnitude dependent on the 
volume fraction of the SPM particles in the bead.

The presence of $H$ alters Eq.~(\ref{rateconst}). Neglecting terms of order $H^{2}/K^{2}$ due to 
the assumption of large anisotropy, yields the following generalized rate constants \cite{aahar}:
\begin{eqnarray}
\label{re1}
\lambda_{0\rightarrow\pi} &=& \lambda_{0}\mbox{exp} (-\frac{V(K+H\mu)}{k_{B}T}),\\
\label{re2}
\lambda_{\pi\rightarrow0} &=& \lambda_{0}\mbox{exp} (-\frac{V(K-H\mu)}{k_{B}T}).
\end{eqnarray}
The relative populations of magnetic moments aligned along and against the anisotropy axis (corresponding to 
$\Phi = 0$ and $\pi$) are now described by a master equation for a two-state system \cite{schakra}:
\begin{equation}
\label{nzero}
\frac{d}{dt}n_{0}(t) = -\lambda_{0\rightarrow\pi} n_{0}(t) 
				+ \lambda_{\pi\rightarrow0}n_{\pi}(t),
\end{equation}
where $n_{0}$ and $n_{\pi}$ denote the fraction of particles with orientation $0$ and $\pi$ respectively. Solving 
Eq.~(\ref{nzero}), we can obtain the time-dependent magnetization
\begin{equation}
\label{mt}
m(t) = V\mu [n_{0}(t)-n_{\pi}(t)]. 
\end{equation}

We now subject the bead to a time-dependent oscillating field $h(t)$ $=$ $h_{o}$$\mbox{cos} \ \small{2}\pi f t$. 
It is necessary to replace the dipolar mean-field $H$ in the exponent of Eqs.~(\ref{re1}) and (\ref{re2}) by a 
time-dependent field $\tilde{H}(t) = H + h(t)$. The hysteresis loop can then be obtained by evaluation of $m(t)$ 
using Eq.~(\ref{mt}) at each value of the applied field $h(t)$. We allow $m(t)$ to evolve for a few cycles of the 
magnetic field to provide sufficient time for the transients to settle down. As the dipolar mean field $H$ is 
weak, the magnetization reversal is primarily driven by the applied field $h(t)$. However we shall see that 
though weak in strength, it significantly affects the relaxation behavior of the magnetic bead. 

It is appropriate to keep a note of the parameter values that are relevent in the context of local heating of 
cells using magnetic nanoparticles. The temperatures required in hyperthermia and chemotherapy treatments are 
usually in the range of 42$^{o}C$ to 45$^{o}C$ \cite{qpankh}. The frequency $f$ of the applied oscillating 
magnetic field is in the 50 to 1500 $KHz$ range and the field amplitude usually varies between 1 to 200 
$Oe$ \cite{qpankh}. Several groups have reported that exposure to fields with a product $h_{o}\cdot f$ not 
exceeding 4.85$\times 10^{8} A \ m^{-1} \ s^{-1}$ is clinically safe \cite{qpankh}. We have ensured that our 
evaluations include the aformentioned ranges of frequency and field amplitude. Further, the problem of tissue 
cooling due to presence of blood flow is difficult to address due to its mathematical complexity. However an 
often-used thumb rule is that a heat deposition rate $P$ of 100 $mW \ cm^{-3}$ is sufficient in most situations
\cite{qpankh}.

Commercially available beads contain nanoparticles of iron, nickel or cobalt compounds with 
tailored volume fractions. Our numerical evaluations of the loop and its area have been made for beads 
comprising of magnetite ($Fe_{3}O_{4}$) nanoparticles having a diameter of 8 $nm$ and an anisotropy constant
of 4.68$\times 10^{5}$ $ergs/cm^{3}$ \cite{gfgoya}. A 1$\%$ volume fraction in this case corresponds to the 
presence of approximately $10^{4}$ magnetite nanoparticles in the micron-sized bead. For body temperature 
$T$ $=$ 37$^{o}C$,  the Ising limit $VK \gg k_{B}T$ that we assume holds for diameters in the range 6 to 
12 $nm$. We find that our qualitative observations remain unchanged, even in the case of MNP of alternative 
compounds, as long as we work in this limit.  

In Figure 1, we plot the response of the bead to the applied field $h(t)$ for several values of the volume 
fraction $V_{f}$. (i) At low volume fractions ($V_{f}\sim 1\%$), the supermoments 
continue to remain non-interacting and the reversal is governed by the Neel relaxation time. The magnetization 
curve is thus the Langevin function. (ii) As the volume fraction is increased, the supermoments begin to interact 
via dipole-dipole coupling. Our numerics indicate that that the resulting mean field $H$ is small for these volume 
fractions. As a consequence the  rates $\lambda_{0\rightarrow\pi}$ and $\lambda_{\pi\rightarrow0}$ governed 
by Eqs.~(\ref{re1}) and (\ref{re2}) respectively have comparable magnitudes and both contribute to the 
relaxation dynamics, especially near $h(t)\approx 0$. The approach to an all-up or an all-down configuration 
of the supermoments is thereby impeded, resulting in smooth and continuous hysteresis loops. (iii) For high 
volume fractions ($V_{f}\gtrsim 30\%$), the mean field $H$ is large enough to make one of the relaxation rates 
dominate over the other for any value of $h(t)$. The system then exhibits a sharp reversal from an all-up to 
an all-down configuration and vice-versa under the action of the oscillating field $h(t)$.

Figure 2a shows the effect of the field frequency $f$ on the hysteresis loop. Leaf-shaped loops are obtained when 
$f \sim \lambda$. These go over to elliptic loops with their major axis parallel to the $h$ axis on increasing 
$f$. For $f \gg \lambda$, the supermoments are unable to respond to the field variation and no 
hysteresis loops are observed. Figure 2b shows the effect of the amplitude of the oscillating field $h_{o}$ on the 
loops. Their shape ceases to depend on $h_{o}$ when it is sufficiently large to produce saturated loops. 

In Figure 3, we quantify the effects of $V_{f}$, $f$ and $h_{o}$ on the loop area in the form
of a scaling law. Our numerical evaluations suggest a power law variation
$A(V_{f},f,h_{o}) \sim V_{f}^{\alpha}f^{\beta} h_{o}^{\delta}$ where the exponents $\alpha$, $\beta$
and $\delta$ are distinct for cases (ii) and (iii) discussed in the context of Figure 1. The values of these
exponents are 1.3$\pm$0.02, 0.95$\pm$0.02 and 1.5$\pm$0.02 for case (ii) and 2.0$\pm$0.02, 0.9$\pm$0.02 
and 1.1$\pm$0.02 for case (iii). The heat deposition rate $P$ can be obtained from the loop area $A$ via the 
relation $P$ $=$ $f\times A/V$ where $V$ is the volume of the bead. This relationship enables us to identify the 
clinically useful range of the loop area which is indicated in the figure. The scaling law can therefore 
be utilized to select (several) triplets $(V_{f},f,h_{o})$ to generate a desired heat deposition rate. 
Some of them, for the encircled values of $A$ which yield $P$ $\sim$ 100 $mW cm^{-3}$, have been provided in Table I.

Finally, we study the phenomenon of return point memory in the bead. If the field $h(t)$ is made to cycle
with a reduced amplitute to generate a sub-loop, the system returns precisely to the same state from which it
left the outer loop. This same memory effect extends to sub-cycles within cycles. The system thus remembers a 
heirarchy of states in its past external fields. Figure 4 illustrates the return point memory effect in the 
magnetic bead. The inset evaluates the heat dissipated as a function of $h_{o}$, the maximal field amplitude
used to generate the minor loops. We find that return point memory is always observed in standard leaf-shaped 
loops with saturation. This phenomenon, though unexplored in the context of hyperthermia, seems a promising 
technique for gradual heating of a local area.

In conclusion, the present letter elucidates hysteresis in a micron-sized bead described as a non-magnetic 
matrix embedded with superparamagnetic nanoparticles. The study enables manipulation of heat dissipation by a 
compatible selection of commercially available beads characterized by the volume fraction $V_{f}$ of the 
superparamagnetic nanoparticles and the frequency $f$ and amplitude $h_{o}$ of the applied oscillating field 
in the laboratory. We hope that the simple mean field framework proves useful in efficient design of 
experiments in the context of therapeutic applications such as hyperthermia and chemotherapy. We also hope 
that this study provokes experiments on the utility of return point memory in these applications.

\vspace{2cm}
\begin{center}
{\bf Table I}\\

\vspace{0.5cm}
\begin{tabular}{|c|c|c|}\hline\hline
$V_{f}$ $(\% )$  & $h_{o}$ (Oe) & $f $ (KHz)\\  \hline
 10  & $ 200 $& $ 1500 - 1800$  \\ \hline
 15  & $ 150 $& $ 1500 - 1800$  \\ \hline
 15  & $ 200 $& $  800 - 1500$  \\ \hline
 20  & $ 100 $& $  900 - 1500$  \\ \hline
 20  & $ 150 $& $  600 - 1200$  \\ \hline
 20  & $ 200 $& $  400 -  850$  \\ \hline
 25  & $ 100 $& $  300 -  800$  \\ \hline
 25  & $ 200 $& $  200 -  350$  \\ \hline
 \hline
\end{tabular}
\end{center}
                        
\newpage

\begin {center}
\begin{figure}
\includegraphics[width=12.0cm,height=12cm,angle=0]{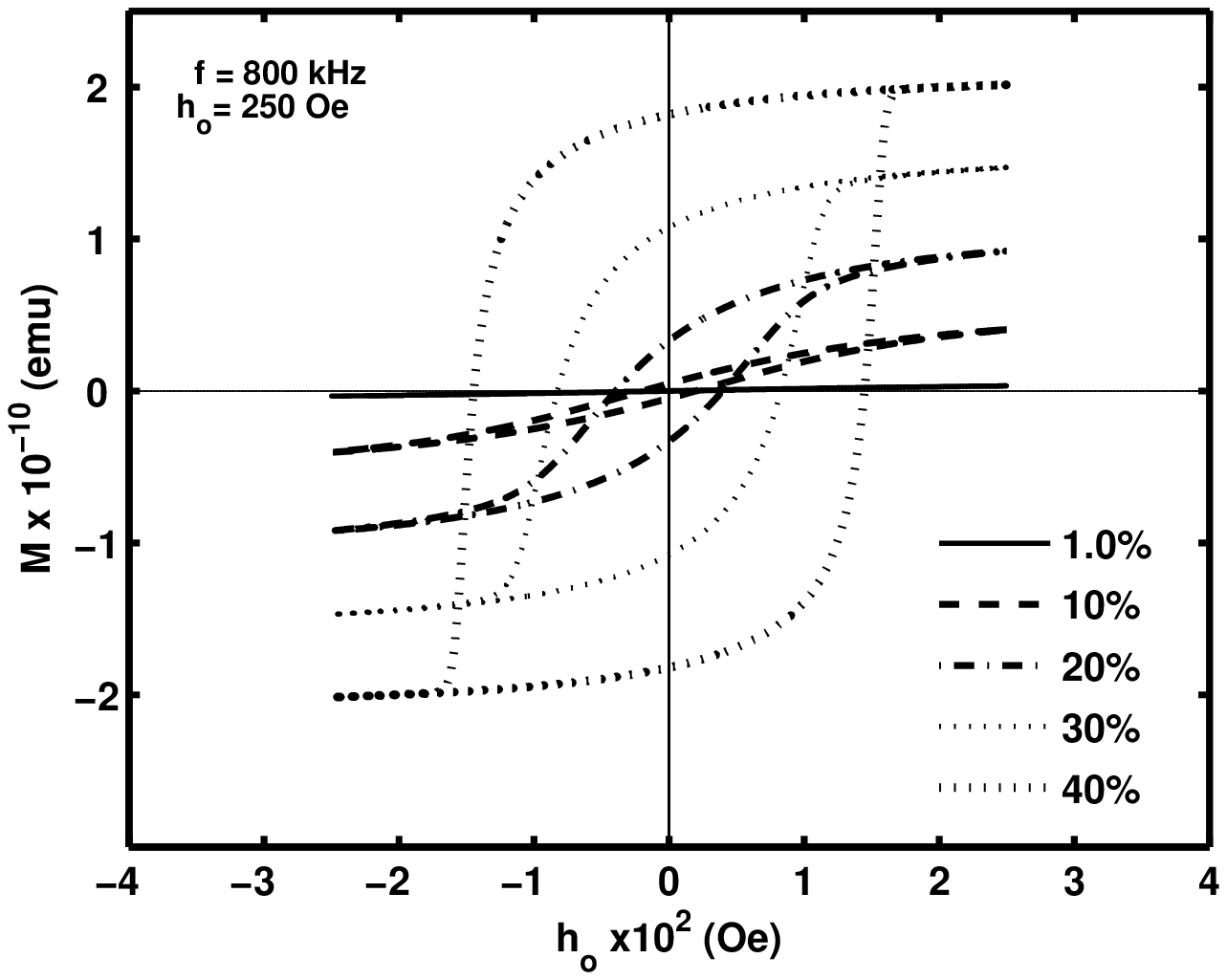}\\
{\bf Figure 1}\\
\end{figure}
\end{center}
\noindent{\bf Figure 1:} Typical hysteresis loops for several values of the volume fraction $V_{f}$ specified
in the figure.   \\
\newpage

\begin{center}
\begin{figure}
\includegraphics[width=12.0cm,height=12cm,angle=0]{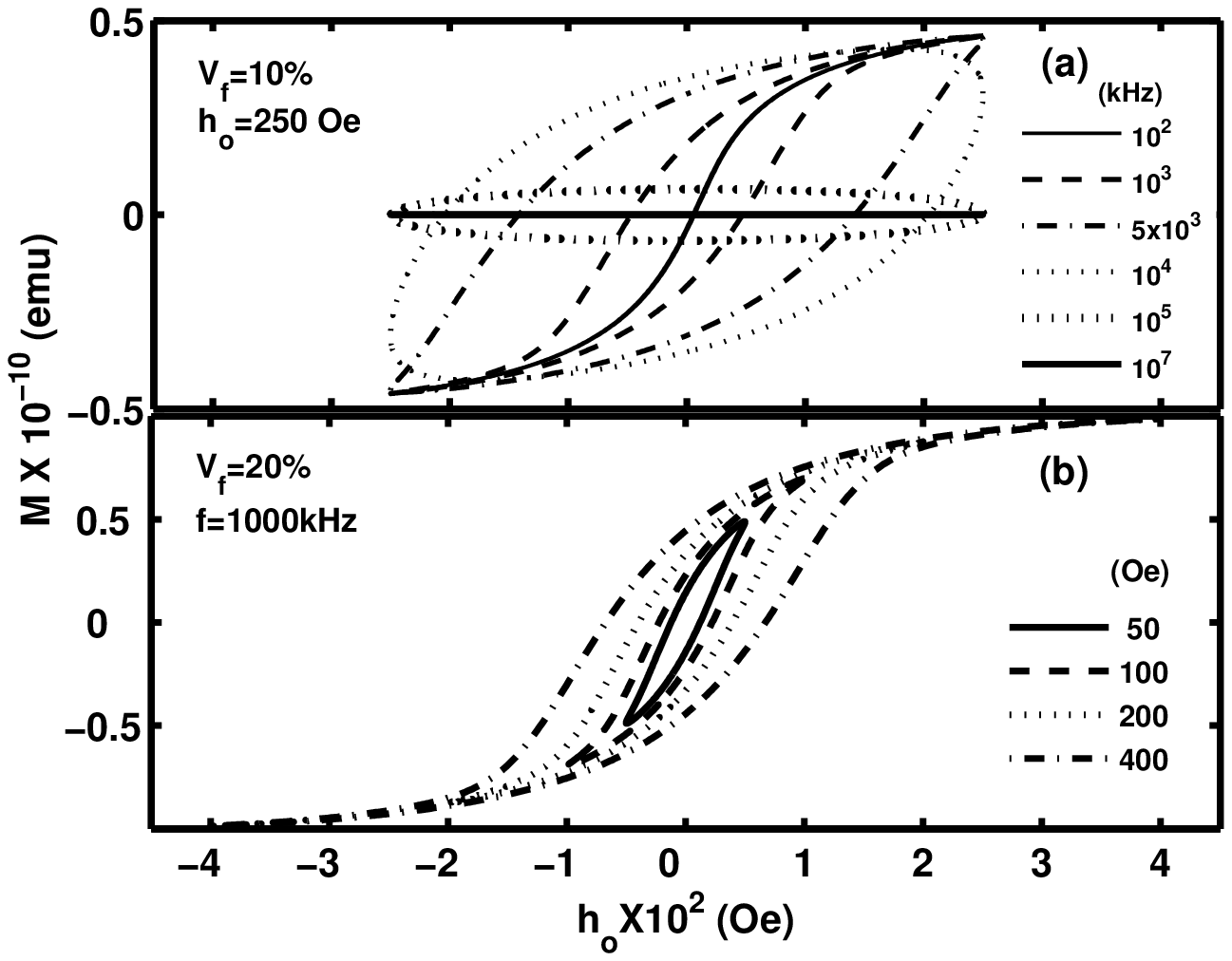}\\
{\bf Figure 2}
\end{figure}
\end{center}
\noindent{\bf Figure 2:} Variation in the hysteresis loop as a function of (a) the applied frequency $f$ and
(b) the amplitude $h_{o}$ of the applied oscillating field.\\

\newpage

\begin {center}
\begin{figure}
\includegraphics[width=12.0cm,height=12cm,angle=0] {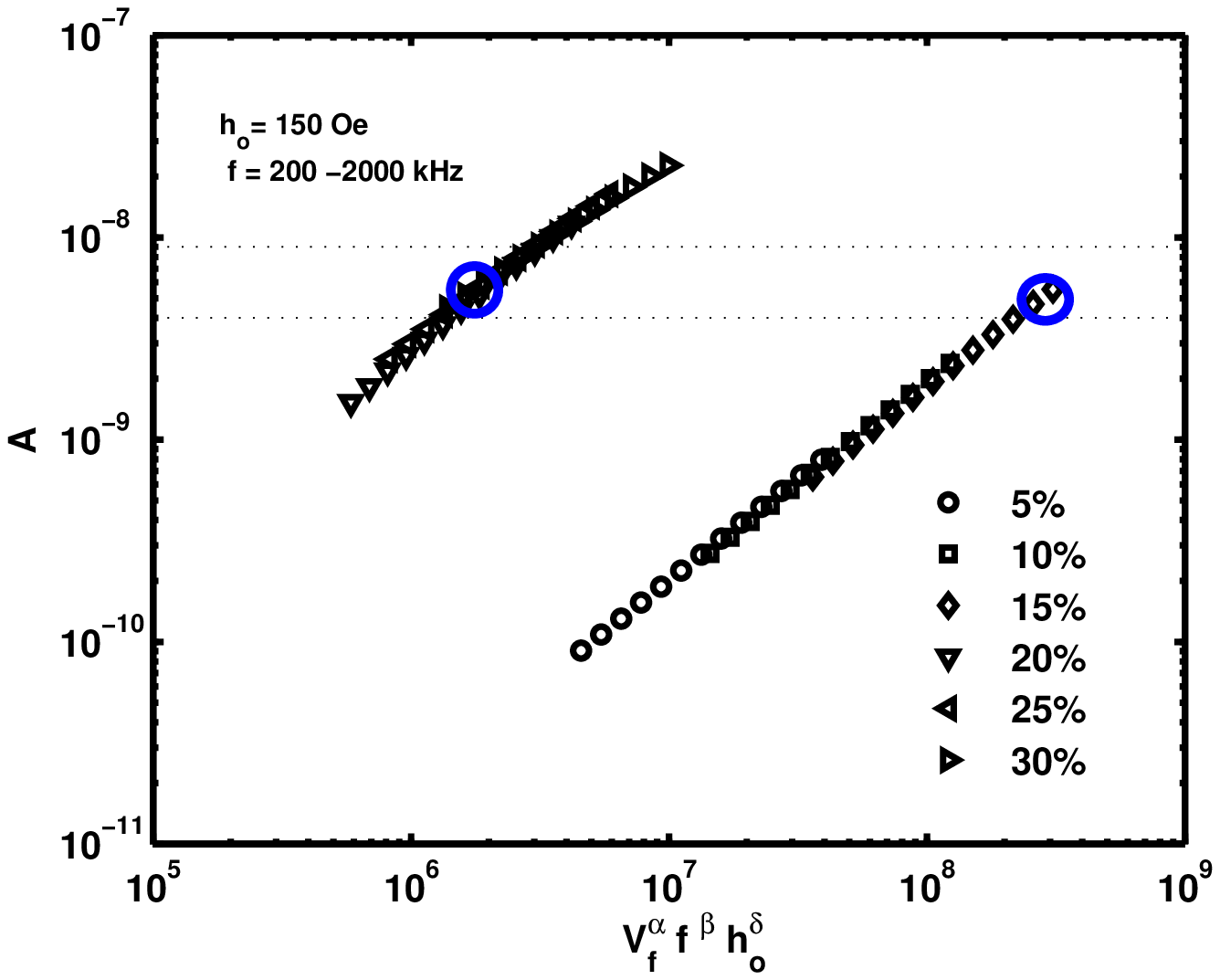}\\
{\bf Figure 3}\\
\end{figure}
\end{center}
\noindent{\bf Figure 3:} A scaling plot which demonstrates that the area of the hysteresis loop scales as 
$A(V_{f},f,h_{o}) \propto V_{f}^{\alpha}f^{\beta} h_{o}^{\delta}$. The scaling exponents are specied in the text.
The region within the dotted lines yields heat deposition suitable for hyperthermia. Several triplets 
($V_{f},f,h_{o}$) corresponding to the encircled value of $A$ are specified in Table I.  
\newpage

\begin{center}
\begin{figure}
\includegraphics[width=12.0cm,height=10cm,angle=0]{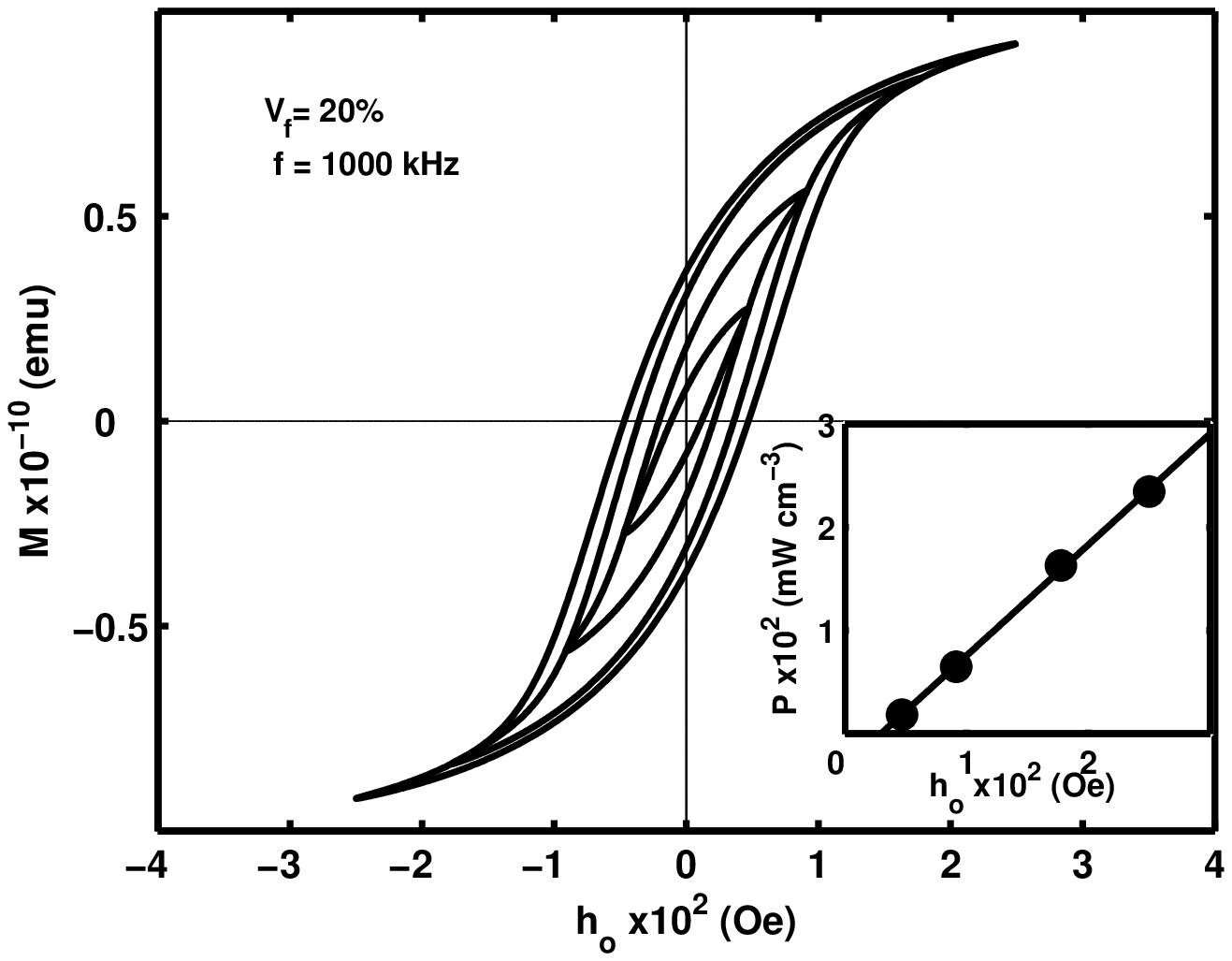}\\
{\bf Figure 4}
\end{figure}
\end{center}
\noindent{\bf Figure 4:} Hysteresis loop showing return point memory for several minor loops. The inset indicates the 
heat dissipated as a function of $h_{o}$, the maximum amplitude of the applied oscillating field for the 
minor loop.

\end{document}